# Robust Input Shaping Vibration Control via Extended Kalman Filter-Incorporated Residual Neural Network

Weiyi Yang, Shuai Li, *Senior Member, IEEE*, Xin Luo, *Senior Member, IEEE*

*Abstract*— With the rapid development of industry, the vibration control of flexible structures and underactuated systems has been increasingly gaining attention. Input shaping technology enables stable performance for high-speed motion in industrial motion systems. However, existing input shapers generally suffer from the ineffective control performance due to the neglect of observation errors. To address this critical issue, this paper proposes an Extended Kalman Filter-incorporated Residual Neural Network-based input Shaping (ERS) model for vibration control. Its main ideas are two-fold: a) adopting an extended Kalman filter to address a vertical flexible beam's model errors; and b) adopting a residual neural network to cascade with the extended Kalman filter for eliminating the remaining observation errors. Detailed experiments on a real dataset collected from a vertical flexible beam demonstrate that the proposed ERS model has achieved significant vibration control performance over several state-of-the-art models.

I. INTRODUCTION

Feedforward control, as a highly effective open-loop control strategy for compensating servo errors, can significantly enhance the system performance [1-7]. As a result, it is extensively utilized in the vibration control for flexible structures and underactuated systems. In the realm of high-speed and high-precision motion control, prevalent feedforward control strategies include input shaping (IS) [8-12], filtering [13], and trajectory smoothing [14-18].

IS is a simple and efficient technique widely applied in vibration suppression for flexible systems. It involves convolving desired input commands with impulse sequences, relying on accurate system parameters rather than detailed system models and motion analysis. Compared to traditional filtering methods such as low-pass filtering [19-22] and notch filtering [23-25], it demonstrates superior vibration suppression capabilities. Another commonly used open-loop control method for addressing vibration control issues is trajectory smoothing [26-29]. Unlike IS, which focuses on designing input impulses, trajectory smoothing relies on models and aims to modify the motion trajectory of systems or robots, ensuring smooth and stable trajectories that meet specific requirements [30-32].

Therefore, IS plays a vital role in the vibration suppression of high-speed motion. In order to achieve superior control performance, both model and observation error identification should be attentively considered [33-36]. To date, model errors have been typically identified through methods such as least squares (LS) [37], Gauss-Newton iteration [38], genetic algorithms (GA) [39], and particle swarm optimization (PSO) [40]. These methods are based on empirical analysis grounded in mathematical and statistical theories. While they exhibit strong interpretability and are easily implemented [41-46], they also come with problems such as repetitive iterative processes, complex genetic operations, and susceptibility to local optima, resulting in a decline in identification accuracy [47-50]. Additionally, due to the highly non-linear nature of servo systems, the extended Kalman filter (EKF) algorithm presents significant advantages [51-55]. Huang *et al.* [56] propose an adaptive generalized extended Kalman filter with unknown input, which eliminates the real-time performance limitation, thereby achieving the optimization of the state and input estimates in terms of minimum variance. Although the above methods can improve the control effectiveness to some extent, they neglect the potential observation errors from external systems, thus failing to fully exploit the control efficiency of the input shaper [57-59].

When identifying observation errors, data-driven compensation models can be highly efficient [60]. However, in real-world scenarios, the sources of observation errors are diverse, including factors like link elasticity deformation [61-64], environmental temperature fluctuations [65-68], mechanical deformations [69-72], and so on. Previous studies [73-76] have shown that neural network (NN)-based models possess high adaptability, flexibility, and learning capabilities in such cases. Nevertheless, they all utilize basic multi-layer neural networks, which have inherent limitations in efficiency and precision [77]. He *et al.* [78] devise the shortcut connection to introduce the residual neural network (ResNN), which effectively eliminates the challenges of training process caused by excessive network depth. Besides, a ResNN demonstrates outstanding performance in non-linear complex systems, exhibiting high compatibility with other optimization models when employed in cascade form [79-81]. Motivated by this discovery, this paper proposes an Extended Kalman Filter-incorporated Residual Neural Network-based input Shaping (ERS) model that cascades an EKF with a ResNN to sequentially address a flexible system's

This research is supported by the National Natural Science Foundation of China under grant 62272078 *(Corresponding Author: Xin Luo).*

Weiyi Yang is with the Chongqing Institute of Green and Intelligent Technology, Chinese Academy of Sciences, Chongqing 400714, China, and also with the Chongqing School, University of Chinese Academy of Sciences, Chongqing 400714, China (e-mail: yangweiyi@cigit.ac.cn).

Shuai Li is with the Faculty of Information Technology and Electrical Engineering, University of Oulu, 90570 Oulu, Finland, and also with the VTT-Technology Research Center of Finland, 90570 Oulu, Finland (e-mail: shuai.li@oulu.fi).

Xin Luo is with the College of Computer and Information Science, Southwest University, Chongqing 400715, China (e-mail: luoxin@swu.edu.cn).

model errors and observation errors, thus achieving highly efficient vibration control performance. Overall, this paper makes the following distinctive contributions:

a) A novel ERC model is proposed, which is able to achieve superior control performance owing to its precise compensation and identification of the model and observation errors;
b) Detailed algorithm design and comprehensive complexity analysis for ERS are presented, which provides guidance for researchers and engineers concerning the issue of vibration control.
c) Extensive experiments are conducted to evaluate the proposed ERS model.

A vertical flexible beam is adopted to evaluate the performance of the proposed ERS model. The outstanding results indicate that the proposed ERS model achieves remarkable performance compared to state-of-the-art models in addressing vibration control issues.

## II. PRELIMINARIES: IS THEORY

The input shaper, comprised of a series of pulses [82], effectively reduces vibration by convolving the input impulses. Specifically, its design varies based on the specific systems, leading to differences in the impulse amplitudes and time delays [43]. The impulse shaping process and convolution principle are shown in Fig. 1.

For an undamped or underdamped second-order system with a transfer function $\omega_n^2/(s^2+2\zeta\omega_n s+\omega_n^2)$ [61], the unit impulse input response at time $t_n$ is given as:

$$g(t) = \frac{\omega_n}{\sqrt{1-\zeta^2}} e^{-\zeta\omega_n(t-t_n)} \sin\omega_d(t-t_n). \tag{1}$$

Note that the system response is expressed as the summation of all the impulses [83], which can be represented as:

$$y(t) = \frac{\omega_n}{\sqrt{1-\zeta^2}} e^{-\zeta\omega_n t} \cdot \sqrt{C^2(\zeta,\omega_n) + S^2(\zeta,\omega_n)} \cdot \sin(\omega_d t - \varphi), \tag{2}$$

where $\omega_d$ is the damped frequency, $\zeta$ is the damping ratio and $\omega_d = \omega_n\sqrt{1-\zeta^2}$, $\varphi$, $C(\zeta,\omega_n)$ and $S(\zeta,\omega_n)$ are given as:

$$\varphi = \arctan\frac{C(\zeta,\omega_n)}{S(\zeta,\omega_n)}, \tag{3}$$

$$\begin{cases} C(\zeta,\omega_n) = \sum_{i=1}^{n} A_i e^{\zeta\omega_n t_i} \cos\omega_d t_i \\ S(\zeta,\omega_n) = \sum_{i=1}^{n} A_i e^{\zeta\omega_n t_i} \sin\omega_d t_i \end{cases}, \tag{4}$$

where $A_i$ and $t_i$ are the amplitudes and time locations of an impulse, $N$ is the number of impulses, and a normalization constraint for $A_i$ is given as:

$$\sum_{i=1}^{N} A_i = 1. \tag{5}$$

Define the residual vibration ratio as:

$$V(\zeta,\omega_n) = e^{-\zeta\omega_n t_N}\sqrt{C^2(\zeta,\omega_n) + S^2(\zeta,\omega_n)}. \tag{6}$$

Therefore, eliminating residual vibrations is equivalent to setting each term in (4) to zero. Notably, the combination of $A_i$ and $t_i$ determines the vibration control effectiveness of the input shaper [84]. Specifically, the ZVD shaper requires the derivative of the vibration variation to zero and incorporates (5) as constraints, which can be obtained as:

$$ZVD = \begin{bmatrix} A_i \\ t_i \end{bmatrix} = \begin{bmatrix} \frac{1}{C} & \frac{2K}{C} & \frac{K^2}{C} \\ 0 & \frac{T_d}{2} & T_d \end{bmatrix}, \tag{7}$$

where $K = e^{-\zeta\pi/\sqrt{1-\zeta^2}}$, $C=1+2K+K^2$ and $T_d = 2\pi/\omega_d$ is the damped period of vibration. However, such shaper ignores errors caused by internal variations within the system and external disturbances, resulting in inefficient vibration control [85]. To address this issue, the next section introduces the ERC model.

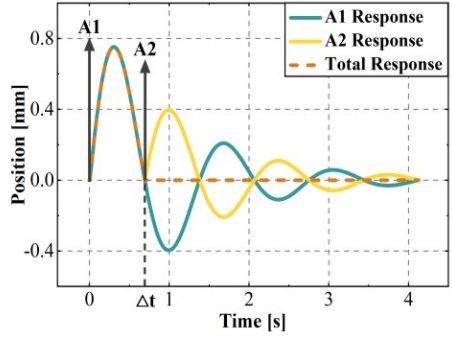

(a) The IS principle. With appropriate amplitude and time delay, impulse A2 can effectively suppress the residual vibrations caused by impulse A1. Note that A2 can be multiple, as long as they can cancel each other out when superimposed.

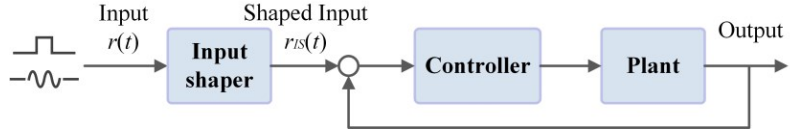

(b) A block diagram of IS control system. Note that in real industrial scenarios, there exist various industrial instruments, whose controllers are rigorously designed and encapsulated, commonly treated as black boxes.

Figure 1. The principle and shaping process of IS.

## III. THE PROPOSED ERS MODEL

### A. The EKF-Based Modeling Parameter Identification

As time passes, variations in the internal structure may result in inaccuracies in the system modeling parameters. To address this issue, we utilize the EKF to compensate the modeling parameters of the flexible system. Hence, using $T$ to represent the shaper's parameters $A_i$ and $t_i$ in Algorithm I, the covariance matrix $P$ is as follows:

$$T_{k|k-1} = T_{k-1|k-1}, \qquad (8)$$

$$P_{k|k-1} = P_{k-1|k-1} + Q_{k-1}, \qquad (9)$$

where $T$ is a length-6 vector consisting of the amplitudes and time locations parameters, $Q_{k-1}$ denotes the covariance matrix of a zero-averaged white noise sequence at the $(k-1)$-th iteration. Defining the measured vibration displacement on the $y$ axis as $\Theta_k$, the optimal Kalman gain $K_k$ and the Jacobian matrix are given as:

$$K_k = P_{k|k-1} J_k^T \left( J_k P_{k|k-1} J_k^T + R_k \right)^{-1}, \qquad (10)$$

$$J = \left[ \frac{\partial \Theta_k}{\partial \omega}, \frac{\partial \Theta_k}{\partial \zeta} \right], \qquad (11)$$

Hence, optimal value for $T$ can be estimated through the computation of the following recursive identification:

$$T_{k|k} = T_{k|k-1} + K_k \left( \Theta_k - J_k T_{k|k-1} \right). \qquad (12)$$

Thus, the covariance matrix $P$ is derived as:

$$P_{k|k} = \left( I - K_k J_k \right) P_{k|k-1}, \qquad (13)$$

where $I$ is the unit matrix. Notably, $k|k-1$ and $k|k$ represent the prior estimate and the posterior estimate at $k$-th iteration, respectively.

Building upon the preceding derivation, the detailed workflow of the modeling parameter identification algorithm, EKF-MPI is introduced, effectively managing errors within the modeling parameters. The subsequent section employs the ResNN to address the remaining observation errors.

### B. The ResNN-Based Observation Error Compensation

As shown in Fig. 1(a), ResNN mainly utilizes the shortcut path principle to mitigate the degradation issue arising from deep network structures [25]. To precisely compensate for observation errors, a 10-layer ResNN is constructed with 40 neurons in each hidden layer and two neurons in the input layer. Specifically, the penultimate layer also consists of two neurons dedicated to learning system parameter biases and the output layer has only one node for estimating the vibration displacement Θ, as shown in Fig. 2(b).

| Algorithm I. EKF-MPI | |
|---|---|
| **Input:** $T_0$, $\{\omega_n, \zeta_n\}$, $\Theta_n$ | |
| **Operation** | **Cost** |
| **Initialize:** $P_0$, $k=1$ | $\Theta(1)$ |
| **Initialize:** $M$: The number of $\Theta_n$ | $\Theta(1)$ |
| set $P_0$ randomly | $\Theta(1)$ |
| set $K_0$ zeros | $\Theta(1)$ |
| **for** $k=1$ **to** $|M|$ | $\times M$ |
|   set $Q_k$ known | $\Theta(1)$ |
|   set $R_k$ known | $\Theta(1)$ |
|   Make $J_k$ evolve with (10). | $\Theta(1)$ |
|   $T_{k|k-1} = T_{k-1|k-1}$ | $\Theta(1)$ |
|   $P_{k|k-1} = P_{k-1|k-1} + Q_{k-1}$ | $\Theta(1)$ |
|   $K_k = P_{k|k-1} JT_k (J_k P_{k|k-1} JT_k + R_k)^{-1}$ | $\Theta(1)$ |
|   $T_{k|k} = T_{k|k-1} + K_k (\Theta_k - J_k T_{k|k-1})$ | $\Theta(1)$ |
|   $P_{k|k} = (I - K_k J_k) P_{k|k-1}$ | $\Theta(1)$ |
|   $k = k+1$ | $\Theta(1)$ |
| **end for** | - |
| $T_{ekf} = T_{k|k}$ | $\Theta(1)$ |
| **Output:** $T_{ekf}$ | |

The activation function $g$ for each neuron in the hidden layers is sigmoid [30], and the input and output of the $i$-th hidden layer are expressed as:

$$h_{i,in} = w_{i,h} h_{i-1,out} + b_{i,h}, \quad h_{i,out} = \left(1 + e^{-h_{i,in}}\right)^{-1}, \qquad (14)$$

where $h_{i,in}$, $h_{i,out}$, $w_{i,h}$ and $b_{i,h}$ denote the input, output, weight and bias of the $i$-th hidden layer, respectively. According to the principles of ResNN, we implement the first residual block between the third and fourth layers. Thus, the input of the fourth layer contains the shortcut output from the second layer and direct output of the third layer:

$$h_{4,in} = w_{4,h} h_{3,out} + b_{4,h} + h_{2,out}, \quad h_{4,out} = \left(1 + e^{-h_{4,in}}\right)^{-1}. \qquad (15)$$

Similarly, the second residual block is established between the sixth and seventh layers and formed as:

$$h_{7,in} = w_{7,h} h_{6,out} + b_{7,h} + h_{5,out}, \quad h_{7,out} = \left(1 + e^{-h_{7,in}}\right)^{-1}. \qquad (16)$$

Finally, the activation function of the output layer is designed as a linear function with a linear coefficient $\alpha$, and its input and output are given as:

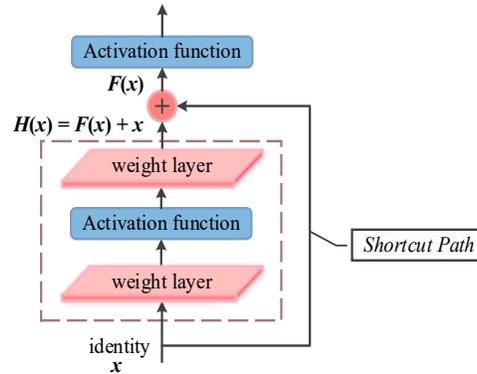

(a) A residual block of shortcut principle, where $x$ and $F(x)$ is the input and residual error calculated through the shortcuts. This skip-layer structure ensures that network performance does not degrade.

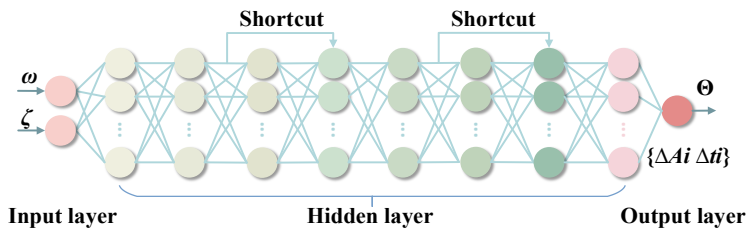

(b) The ResNN structure (2-40-40-40-40-40-40-40-6-1) with shortcuts from the 2nd to the 4th layer and the 5th to the 7th layer, respectively.

Figure 2. The neural network of this paper.

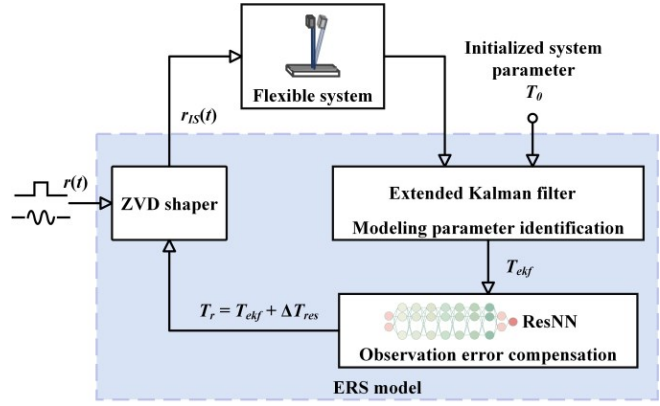

Figure 3. Workflow chart of ERS model.

$$y_{in} = w_o h_{8,out} + b_o, \quad y_{out} = \alpha y_{in}. \quad (17)$$

Subsequently, the error between the actual vibration displacement and the displacement estimated by ResNN is backpropagated [65-69]. Notably, the loss function of the designed network is given as:

$$E = \frac{1}{2} \sum_{i=1}^{n} \|\Theta_i - \theta_i\|^2, \quad (18)$$

where $\Theta_i$ and $\theta_i$ are the measured and nominal vibration displacement of the flexible system, respectively. According to the chain rule [86], the formula for weight updates is as follows:

$$\Delta w_o = \frac{\partial E}{\partial w_o} = \frac{\partial E}{\partial y_{out}} \cdot \frac{\partial y_{out}}{\partial y_{in}} \cdot \frac{\partial y_{in}}{\partial w_o} = \sum_{i=1}^{n}(Y_i - L_i) \cdot h_{8\_out}, \quad (19)$$

Accordingly, the other parameters are updated as:

$$\Delta w_{8\_h} = \sum_{i=1}^{n}(Y_i - L_i) \cdot w \cdot h_{8\_out}(1 - h_{8\_out}) \cdot h_{7\_out}, \quad (20a)$$

$$\Delta w_{7\_h} = \Delta w_{8\_h} \cdot w_{8\_h}(1 - h_{7\_out}) \cdot h_{6\_out}, \quad (20b)$$

$$\Delta w_{6\_h} = \Delta w_{7\_h} \cdot w_{7\_out}(1 - h_{6\_out}) \cdot h_{5\_out}, \quad (20c)$$

$$\Delta w_{5\_h} = \Delta w_{6\_h} \cdot w_{6\_out}(1 - h_{5\_out}) \cdot h_{4\_out}, \quad (20d)$$

$$\Delta w_{4\_h} = \Delta w_{5\_h} \cdot w_{5\_out}(1 - h_{4\_out}) \cdot h_{3\_out}, \quad (20e)$$

$$\Delta w_{3\_h} = \Delta w_{4\_h} \cdot w_{4\_out}(1 - h_{3\_out}) \cdot h_{2\_out}, \quad (20f)$$

$$\Delta w_{2\_h} = \Delta w_{3\_h} \cdot w_{3\_out}(1 - h_{2\_out}) \cdot h_{1\_out}, \quad (20g)$$

$$\Delta w_{1\_h} = \Delta w_{2\_h} \cdot w_{2\_out}(1 - h_{1\_out}) \cdot x_{in}, \quad (20h)$$

$$w_{i\_h} = w_{i\_h} - \rho_i \cdot \Delta w_{i\_h}, \quad (20i)$$

where $\rho$ is a constant and Fig. 3 illustrates the whole training process. Note that the following error function serves as the termination condition for training:

$$\Upsilon = \sum_{i=1}^{n} \|\Theta_i - \theta_i\|. \quad (21)$$

As parameters $\Delta T_{res}$ can be identified by Algorithm II, for each defined nominal system parameters $T_n$, the actual system parameters $T_r$ is given as:

$$T_r = T_n + \Delta T_{res}. \quad (22)$$

As shown in Algorithm II ResNN-OEC, an algorithm for the ResNN-based observation error compensation is presented based on the above inferences.

## IV. EXPERIMENTS RESULTS AND COMPARISON

### A. General Settings

*1) Evaluation Protocol.* We adopt the following four metrics four metrics i.e., maximum transient swing (MTS) [16], maximum error (MAX), mean square error (RMSE) and mean error (MEAN) [8], to evaluate the performance of a tested model:

$$MTS = \max\{|Z_i|\}, \quad (24a)$$

$$MAX = \max|Z_i - \hat{Z}_i|, \quad (24b)$$

$$RMSE = \sqrt{\frac{1}{n}\sum_{i=1}^{n}(Z_i - \hat{Z}_i)^2}, \quad (24c)$$

$$MEAN = \frac{1}{n}\sum_{i=1}^{n}|Z_i - \hat{Z}_i|, \quad (24d)$$

*2) Datasets.* Vibrating flexible beam (VFB): it is collected from a flexible beam perpendicular to the linear. Specifically, we constructed two datasets by varying the payloads and link lengths: VFB-1 (samples:100, payload: 0.6kg, length: 0.3m) and VFB-2 (samples:100, payload: 0.3kg, length: 0.5m).

Note that these samples are in the form of two system parameters (the damped frequency $\omega$ and the damping ratio $\zeta$) and vibration displacements $\Theta$, as shown in Table I. Specifically, the 90%-10% training-testing setting is employed to verify the tested models. The final averaged results along with the standard deviations by each model are carefully recorded for reporting objective results. For each tested model on each data case, the termination condition is set as a) the objective function decreasing tendency between two consecutive training iterations becomes smaller than $10^{-4}$, or b) the training iteration count exceeds 100.

| Algorithm II. ResNN-OEC | |
|---|---|
| **Input:** $\{\omega_n, \zeta_n\}, \Theta_n, T_{ekf}$ | |
| **Operation** | **Cost** |
| **Initialize:** $w_0, L_n, r=0, n=1, R=100$ | $\Theta(1)$ |
| **Initialize:** $N$: The number of $\{\omega_n, \zeta_n\}$ | $\Theta(1)$ |
| **while** not converge && $r \leq R$ **do** | $\times r$ |
|   **for** $n=1$ **to** $|N|$ | $\times N$ |
|     $h_{0, out} = \{\omega_n, \zeta_n\}$ | $\Theta(1)$ |
|     Compute $h_{j,in}$ and $h_{j,out}$ from $j=1$ to 3 with (14) | $\Theta(1)$ |
|     Compute $h_{4, in}$ and $h_{4, out}$ with (15) | $\Theta(1)$ |
|     Compute $h_{j,in}$ and $h_{j,out}$ from $j=5$ to 6 with (14) | $\Theta(1)$ |
|     Compute $h_{7, in}$ and $h_{7, out}$ with (16) | $\Theta(1)$ |
|     Compute $h_{8, in}$ and $h_{8, out}$ with (14) | $\Theta(1)$ |
|     Compute $y_{in}$ and $y_{out}$ with (17) | $\Theta(1)$ |
|     $\Delta T_{res} = y_{out}$ | $\Theta(1)$ |
|     $\Delta E = \Theta_n - \theta_n$ | $\Theta(1)$ |
|     Update $w_0 \sim w_8$ with (19) ~ (20i) | $\Theta(1)$ |
|   **end for** | - |
| **end while** | - |
| $T_r = T_{ekf} + \Delta T_{res}$ | $\Theta(1)$ |
| **Output:** $T_r$ | |

| Algorithm III. ERS | |
|---|---|
| **Input:** $T_0, \{\omega_n, \zeta_n\}, \Theta_n$ | |
| /--Note: *Initialization*--/ | |
| 1 **Initialize:** $M, N, R=100$ | $T_1$ |
| 2 **Initialize:** $L_n, r=0, n=1, X=D_0$ | |
| /--Note: *E-Step*--/ | |
| 3 $D_{ekf}$=**EKF-GPI**$(T_0, \{\omega_n, \zeta_n\}, \Theta_n)$ | $T_2$ |
| /--Note: *R-Step*--/ | |
| 17 $D_f$= **ResNN-OEC**$(T_{ekf}, \{\omega_n, \zeta_n\}, \Theta_n)$ | $T_3$ |
| /--Note: **Training Ends**--/ | |
| **Output:** $T_r$ | |

TABLE I. EXAMPLE OF DATA SAMPLES FROM VBF

| No. | $\omega_n$/Hz | $\zeta$/- | $\Theta$/mm |
|---|---|---|---|
| 1 | 5.00 | 0.10 | 24.90 |
| 2 | 10.00 | 0.20 | 31.18 |
| 3 | 15.00 | 0.30 | 31.28 |

TABLE II. COMPARATIVE TEST ERROR (IN CM) RESULTS

| Dataset | Metric | M1 | M2 | M3 | M4 | M5 | M6 | M7 | M8 |
|---|---|---|---|---|---|---|---|---|---|
| **VFB-1** | MAX | 0.34±0.01 | 0.44±0.02 | 0.21±0.00 | 0.52±0.00 | 0.42+0.00 | 0.26±0.00 | 0.14±0.00* | 2.72±0.06 |
|  | RMSE | 0.39±0.02 | 2.29±0.05 | 0.26±0.00 | 0.57±0.00 | 0.48±0.00 | 0.32±0.01 | 0.16+0.00* | 2.51±0.03 |
|  | MEAN | 0.65±0.03 | 0.86±0.03 | 0.57±0.02 | 0.93±01 | 0.80±0.01 | 0.62±0.02 | 0.27±0.01* | 2.50±0.03 |
| **VFB-2** | MAX | 0.92±0.33 | 0.90±0.49 | 0.78±0.37 | 1.19±0.39 | 1.16±0.48 | 0.86±0.25 | 0.48±0.06* | 3.20±0.07 |
|  | RMSE | 0.40±0.15 | 2.83±0.64 | 0.31±0.14 | 0.60±0.21 | 0.56±0.24 | 0.39±0.74 | 0.27±0.03* | 3.30±0.06 |
|  | MEAN | 0.29±0.13 | 0.36±0.19 | 0.25±0.08 | 0.50±0.20 | 0.44±0.20 | 0.30±0.73 | 0.23±0.04* | 3.60±0.02 |

Values marked with * indicate the best performing data in the comparison experiment.
M8 stands for the results before parameter identification, i.e., unoptimized ZVD shaper.

TABLE III. TIME COST OF METHODS M4-M10 ON RMSE

| Dataset | Item | M1 | M2 | M3 | M4 | M5 | M6 | M7 |
|---|---|---|---|---|---|---|---|---|
| **VFB-1** | Iteration | 6 | 15 | 10 | 14 | 25 | 17 | 4* |
|  | Time (s) | 2.07±0.40E-2 | 5.67±0.08E-2 | 0.86±0.13E-2* | 0.98±0.03E-2 | 1.81±0.07E-1 | 4.08±0.09E-2 | 2.24±0.06E-2 |
| **VFB-2** | Iteration | 4 | 18 | 3* | 12 | 15 | 11 | 13 |
|  | Time (s) | 1.91±0.32E-2 | 1.68±0.12E-1 | 0.61±0.36E-2* | 2.59±0.17E-2 | 3.39±0.32E-1 | 7.65±0.28E-2 | 3.99±0.06E-2 |

Values marked with * indicate the best performing data in the comparison experiment.

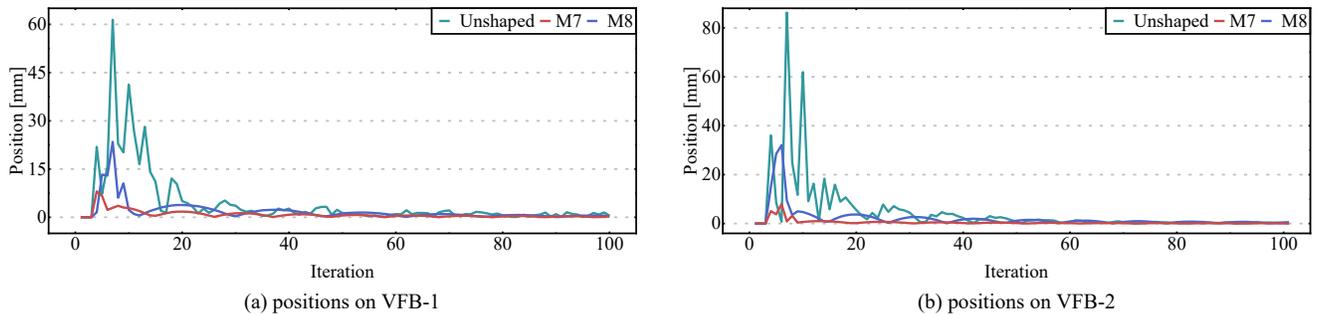

(a) positions on VFB-1   (b) positions on VFB-2

Figure 4. Comparison of actual vibration positions on VFB-1 and VFB-2.

### B. Performance Comparison

**Compared Models.** This part of experiments compares an EKF-ResNN based ZVD shaper with several state-of-the-art input shapers for performance validation. The involved models are listed below:
a) **M1**: An LM [47]-based ZVD shaper.
b) **M2**: A PSO [38]-based ZVD shaper.
c) **M3**: An EKF [41]-based ZVD shaper.
d) **M4**: A single-layered NN [60]-based ZVD shaper. Differently from the ResNN designed in this article, it consists of only two hidden layers with 40 and 2 neurons, respectively.
e) **M5**: A multi-layered NN [60]-based ZVD shaper, whose network structure is 2-40-40-40-40-40-40-40-2-1.
f) **M6**: An ResNN [61] with ZVD shaper.
g) **M7**: The ERS presented in the article.
h) **M8**: An unoptimized ZVD [16] shaper.

Table II lists the vibration control results of the comparative experiment.

The proposed ERS model outperforms state-of-the-art identification models. As shown in Table II, on VFB-1, the MAX error of M7 is 0.14 mm, which is the smallest error, 94.85% less than that of M8. Similar results are encountered with VFB-2. Moreover, as depicted in Fig. 4, we utilize the most effective parameters identified from M7 in two practical systems, resulting in a significant vibration suppression effect.

TABLE IV. PARAMETERS AFTER IDENTIFICATION ON THE TWO DATASETS

| Setup | $A_1$/- | $A_2$/- | $A_3$/- | $t_1$/s | $t_2$/s | $t_3$/s |
|---|---|---|---|---|---|---|
| VFB-1 | 0.01 | 0.02 | 0.24 | 0.14 | 0.43 | 0.44 |
| VFB-2 | 0.01 | 0.01 | 0.01 | 0.03 | 0.04 | 0.07 |

M7 is trained sequentially, therefore, its time cost is the sum of the training times of its two components. Nonetheless, even with this consideration, the training time of M7 is faster than some models. Specifically, on VFB-1, it is 60.49%, 87.62%, and 45.09% faster than M2, M5, and M6, respectively. On VFB-2, it is 76.25%, 88.23%, and 47.84% faster than M2, M5, and M6, respectively.

## V. Conclusion

This paper proposes an ERS model to achieve highly accurate vibration control for flexible systems. By effectively cascading model parameter identification based on EKF and observation parameter compensation based on ResNN, the ERS model demonstrates higher control accuracy compared to the state-of-the-art models. However, due to its serial nature, it also brings the drawback of longer time costs. Therefore, it is well-suited for industrial applications aiming for the utmost precision in control. In the future, we plan to investigate acceleration methods for ERS models based on GPU or other parallelization technologies, exploring the potential of online control.

## References

[1] S.M. Fasih ur Rehman, Z. Mohamed, A.R. Husain, H.I. Jaafar, M.H. Shaheed and M.A. Abbasi, "Input shaping with an adaptive scheme for swing control of an underactuated tower crane under payload hoisting and mass variations," *Mechanical Systems and Signal Processing*, vol. 175, no. 1, pp. 109106, Aug. 2022.

[2] X. Liao, K. Hoang, and X. Luo, "Local Search-based Anytime Algorithms for Continuous Distributed Constraint Optimization Problems," *IEEE/CAA Journal of Automatica Sinica*, DOI: 10.1109/JAS.2024.124413.

[3] Y. Zhong and X. Luo, "Alternating-direction-method of Multipliers-Based Symmetric Nonnegative Latent Factor Analysis for Large-scale Undirected Weighted Networks," in *Proc. of 2021 IEEE 17th International Conference on Automation Science and Engineering*, Lyon, France, 2021, pp. 1527-1532.

[4] J. Chen, Y. Yuan and X. Luo, "SDGNN: Symmetry-Preserving Dual-Stream Graph Neural Networks," *IEEE/CAA Journal of Automatica Sinica*, vol. 11, no. 7, pp. 1717-1719, July 2024.

[5] W. Qin and X. Luo, "Asynchronous Parallel Fuzzy Stochastic Gradient Descent for High-Dimensional Incomplete Data Representation," *IEEE Transactions on Fuzzy Systems*, vol. 32, no. 2, pp. 445-459, Feb. 2024.

[6] Q. Jiang et al., "Iterative Role Negotiation via the Bilevel GRA++ With Decision Tolerance," *IEEE Transactions on Computational Social Systems*, DOI: 10.1109/TCSS.2024.3409893.

[7] Y. Xia, K. Xiao, Y. Yao, Z. Geng and Z. Lendek, 'Fixed-Time Fuzzy Vibration Reduction for Stochastic MEMS Gyroscopes with Low Communication Resources," *IEEE Transactions on Fuzzy Systems*, DOI: 10.1109/TFUZZ.2024.3392296.

[8] W. Yang, S. Li, Z. Li and X. Luo, "Highly-Accurate Manipulator Calibration via Extended Kalman Filter-Incorporated Residual Neural Network," *IEEE Transactions on Industrial Informatics*, vol. 19, no. 11, pp. 10831-10841, Nov. 2023.

[9] J. Li et al., "Saliency-Aware Dual Embedded Attention Network for Multivariate Time-Series Forecasting in Information Technology Operations," *IEEE Transactions on Industrial Informatics*, vol. 20, no. 3, pp. 4206-4217, March 2024.

[10] N. Zeng, X. Li, P. Wu, H. Li and X. Luo, "A Novel Tensor Decomposition-Based Efficient Detector for Low-Altitude Aerial Objects With Knowledge Distillation Scheme," *IEEE/CAA Journal of Automatica Sinica*, vol. 11, no. 2, pp. 487-501, February 2024.

[11] X. Luo, L. Wang, P. Hu and L. Hu, "Predicting Protein-Protein Interactions Using Sequence and Network Information via Variational Graph Autoencoder," *IEEE/ACM Transactions on Computational Biology and Bioinformatics*, vol. 20, no. 5, pp. 3182-3194, 1 Sept.-Oct. 2023.

[12] L. Hu, Y. Yang, Z. Tang, Y. He and X. Luo, "FCAN-MOPSO: An Improved Fuzzy-Based Graph Clustering Algorithm for Complex Networks With Multiobjective Particle Swarm Optimization," *IEEE Transactions on Fuzzy Systems*, vol. 31, no. 10, pp. 3470-3484, Oct. 2023.

[13] Y. Sun et al., "Underwater vibration adhesion by frequency-controlled rigid disc for underwater robotics grasping," *IEEE Robotics and Automation Letters*, DOI: 10.1109/LRA.2024.3392438.

[14] Z. Zhang, Y. Guo, D. Gong and J. Liu, "Global Integral Sliding-Mode Control With Improved Nonlinear Extended State Observer for Rotary Tracking of a Hydraulic Roofbolter," *IEEE/ASME Transactions on Mechatronics*, vol. 28, no. 1, pp. 483-494, Feb. 2023.

[15] D. Wu, Y. He and X. Luo, "A Graph-Incorporated Latent Factor Analysis Model for High-Dimensional and Sparse Data," *IEEE Transactions on Emerging Topics in Computing*, vol. 11, no. 4, pp. 907-917, Oct.-Dec. 2023.

[16] A. A. M. Awi, S. S. N. N. Zawawi, L. Ramli and I. M. Lazim, "Robust Input Shaping for Swing Control of an Overhead Crane," *Asia Simulation Conference*, vol. 1912, no. 1, pp. 180-187, Oct. 2023.

[17] Z. Li, S. Li, O. O. Bamasag, A. Alhothali and X. Luo, "Diversified Regularization Enhanced Training for Effective Manipulator Calibration," *IEEE Transactions on Neural Networks and Learning Systems*, vol. 34, no. 11, pp. 8778-8790, Nov. 2023.

[18] Y. Zhou, X. Luo and M. Zhou, "Cryptocurrency Transaction Network Embedding From Static and Dynamic Perspectives: An Overview," *IEEE/CAA Journal of Automatica Sinica*, vol. 10, no. 5, pp. 1105-1121, May. 2023.

[19] R. Rehammar and S. Gasparinetti, "Low-Pass Filter With Ultrawide Stopband for Quantum Computing Applications," *IEEE Transactions on Microwave Theory and Techniques*, vol. 71, no. 7, pp. 3075-3080, Jul. 2023.

[20] Z. Liu, Y. Yi and X. Luo, "A High-Order Proximity-Incorporated Nonnegative Matrix Factorization-Based Community Detector," *IEEE Transactions on Emerging Topics in Computational Intelligence*, vol. 7, no. 3, pp. 700-714, June 2023.

[21] Z. Xie, L. Jin and X. Luo, "Kinematics-Based Motion-Force Control for Redundant Manipulators With Quaternion Control," *IEEE Transactions on Automation Science and Engineering*, vol. 20, no. 3, pp. 1815-1828, July 2023.

[22] M. Chen, C. He and X. Luo, "MNL: A Highly-Efficient Model for Large-scale Dynamic Weighted Directed Network Representation," *IEEE Transactions on Big Data*, vol. 9, no. 3, pp. 889-903, 1 June 2023.

[23] Y. Yao, D. Xu, Y. Chen, F. Peng and Y. Huang, "Robust Notch Filter-Based Active Damping Design for LCL-Equipped High-Speed PMSMs Considering Dual Resonance Problem," *IEEE Transactions on Industrial Electronics*, DOI: 10.1109/TIE.2024.3368157.

[24] W. Li, R. Wang, X. Luo and M. Zhou, "A Second-Order Symmetric Non-Negative Latent Factor Model for Undirected Weighted Network Representation," *IEEE Transactions on Network Science and Engineering*, vol. 10, no. 2, pp. 606-618, 1 March-April 2023.

[25] X. Xu, M. Lin, X. Luo and Z. Xu, "HRST-LR: A Hessian Regularization Spatio-Temporal Low Rank Algorithm for Traffic Data Imputation," *IEEE Transactions on Intelligent Transportation Systems*, vol. 24, no. 10, pp. 11001-11017, Oct. 2023.

[26] Y. Liu and Z. Yang, "Trajectory Smoothing Algorithm Based on Kalman Filter," in *Proc. of 2023 7th International Conference on Machine Vision and Information Technology*, Xiamen, China, Mar. 2023, pp. 52-56.

[27] X. Luo, H. Wu, Z. Wang, J. Wang and D. Meng, "A Novel Approach to Large-Scale Dynamically Weighted Directed Network Representation," I*EEE Transactions on Pattern Analysis and Machine Intelligence*, vol. 44, no. 12, pp. 9756-9773, 1 Dec. 2022.

[28] X. Luo, Y. Yuan, S. Chen, N. Zeng and Z. Wang, "Position-Transitional Particle Swarm Optimization-Incorporated Latent Factor Analysis," *IEEE Transactions on Knowledge and Data Engineering*, vol. 34, no. 8, pp. 3958-3970, 1 Aug. 2022.

[29] D. Wu, X. Luo, M. Shang, Y. He, G. Wang and X. Wu, "A Data-Characteristic-Aware Latent Factor Model for Web Services QoS Prediction," *IEEE Transactions on Knowledge and Data Engineering*, vol. 34, no. 6, pp. 2525-2538, 1 June 2022.

[30] T. Bloemers, S. Leemrijse, V. Preda, F. Boquet, T. Oomen and R. Tóth, "Vibration Control Under Frequency-Varying Disturbances With Application to Satellites," *IEEE Transactions on Control Systems Technology*, DOI: 10.1109/TCST.2024.3384896.

[31] X. Luo, H. Wu and Z. Li, "Neulft: A Novel Approach to Nonlinear Canonical Polyadic Decomposition on High-Dimensional Incomplete Tensors," *IEEE Transactions on Knowledge and Data Engineering*, vol. 35, no. 6, pp. 6148-6166, 1 June 2023.

[32] X. Luo, Y. Zhou, Z. Liu and M. Zhou, "Fast and Accurate Non-Negative Latent Factor Analysis of High-Dimensional and Sparse Matrices in Recommender Systems," *IEEE Transactions on Knowledge and Data Engineering*, vol. 35, no. 4, pp. 3897-3911, 1 April 2023.

[33] W. Yang, S. Li and X. Luo, "Data Driven Vibration Control: A Review," *IEEE/CAA Journal of Automatica Sinica*, DOI: 10.1109/JAS.2024.124431.

[34] D. Wu, P. Zhang, Y. He and X. Luo, "A Double-Space and Double-Norm Ensembled Latent Factor Model for Highly Accurate Web Service QoS Prediction," *IEEE Transactions on Services Computing*, vol. 16, no. 2, pp. 802-814, 1 March-April 2023.

[35] D. Wu, X. Luo, Y. He and M. Zhou, "A Prediction-Sampling-Based Multilayer-Structured Latent Factor Model for Accurate Representation to High-Dimensional and Sparse Data," *IEEE Transactions on Neural Networks and Learning Systems*, vol. 35, no. 3, pp. 3845-3858, March 2024.

[36] H. Wu, X. Luo, M. Zhou, M. J. Rawa, K. Sedraoui and A. Albeshri, "A PID-incorporated Latent Factorization of Tensors Approach to Dynamically Weighted Directed Network Analysis," *IEEE/CAA Journal of Automatica Sinica*, vol. 9, no. 3, pp. 533-546, March 2022.

[37] L. Li, Q. Zhang, T. Zhang and Y. Zou, "Vibration suppression of ball-screw drive system based on flexible dynamics model," *Engineering Applications of Artificial Intelligence*, vol. 117, no. 1, pp. 105506, Jan. 2023.

[38] W. Wang, C. Hu, K. Zhou and Z. Wang, "Time Parameter Mapping and Contour Error Precompensation for Multiaxis Input Shaping," *IEEE Transactions on Industrial Informatics*, vol. 19, no. 3, pp. 2640-2651, Mar. 2023.

[39] T. Yi, Q. Pei, D. Li, S. Wei, D. Jia and H. Zhang, "Optimization of Simulation Parameters of Input Shaper Based on Genetic Algorithm," in *Proc. of 2022 Second International Conference on Advanced Technologies in Intelligent Control, Environment, Computing & Communication Engineering*, Bangalore, India, Dec. 2022, pp. 1-6.

[40] B. Xu, R. Wang, B. Peng, F. A. Alqurashi and M. Salama, "Automatic parameter selection ZVD shaping algorithm for crane vibration suppression based on particle swarm optimization," *Applied Mathematics and Nonlinear Sciences*, vol. 7, no. 1, pp. 73-82, Jan. 2022.

[41] G. Chen, L. Zhu, W. Zheng and C. Wang, "Vibration Suppression in a Two-Mass Drive System Using Three Input Shaping - Comparative Study," in *Proc. of 2023 International Conference on Computing, Electronics & Communications Engineering*, Swansea, United Kingdom, Sep. 2023, pp. 68-73.

[42] L. Hu, S. Yang, X. Luo and M. Zhou, "An Algorithm of Inductively Identifying Clusters From Attributed Graphs," *IEEE Transactions on Big Data*, vol. 8, no. 2, pp. 523-534, 1 April 2022.

[43] L. Hu, S. Yang, X. Luo and M. Zhou, "An Algorithm of Inductively Identifying Clusters From Attributed Graphs," *IEEE Transactions on Big Data*, vol. 8, no. 2, pp. 523-534, 1 April 2022.

[44] X. Luo, M. Chen, H. Wu, Z. Liu, H. Yuan and M. Zhou, "Adjusting Learning Depth in Nonnegative Latent Factorization of Tensors for Accurately Modeling Temporal Patterns in Dynamic QoS Data," *IEEE Transactions on Automation Science and Engineering*, vol. 18, no. 4, pp. 2142-2155, Oct. 2021.

[45] L. Hu, J. Zhang, X. Pan, X. Luo and H. Yuan, "An Effective Link-Based Clustering Algorithm for Detecting Overlapping Protein Complexes in Protein-Protein Interaction Networks," *IEEE Transactions on Network Science and Engineering*, vol. 8, no. 4, pp. 3275-3289, 1 Oct.-Dec. 2021.

[46] Y. Yuan, X. Luo, M. Shang and Z. Wang, "A Kalman-Filter-Incorporated Latent Factor Analysis Model for Temporally Dynamic Sparse Data," *IEEE Transactions on Cybernetics*, vol. 53, no. 9, pp. 5788-5801, Sept. 2023.

[47] W. Tang, R. Ma, W. Wang, T. Xu, H. Gao and X. Wang, "Composite Control of Overhead Cranes Based on Input Shaper and PID," in *Proc. of 2023 42nd Chinese Control Conference*, Tianjin, China, Jul. 2023, pp. 2568-2572.

[48] X. Luo, Y. Zhong, Z. Wang and M. Li, "An Alternating-Direction-Method of Multipliers-Incorporated Approach to Symmetric Non-Negative Latent Factor Analysis," *IEEE Transactions on Neural Networks and Learning Systems*, vol. 34, no. 8, pp. 4826-4840, Aug. 2023.

[49] A. Gavula, P. Hubinský and A. Babinec, "Damping of Oscillations of a Rotary Pendulum System," *Applied Sciences*, vol. 13, no. 21, pp. 11946, Nov. 2023.

[50] Z. Li, S. Li and X. Luo, "Efficient Industrial Robot Calibration via a Novel Unscented Kalman Filter-Incorporated Variable Step-Size Levenberg–Marquardt Algorithm," *IEEE Transactions on Instrumentation and Measurement*, vol. 72, pp. 1-12, Apr. 2023.

[51] Y. Deng, X. Hou, B. Li, J. Wang and Y. Zhang. "A highly powerful calibration method for robotic smoothing system calibration via using adaptive residual extended Kalman filter". *Robotics and Computer-Integrated Manufacturing*, vol. 86, pp. 102660, Apr. 2024.

[52] W. Li, X. Luo, H. Yuan and M. Zhou, "A Momentum-Accelerated Hessian-Vector-Based Latent Factor Analysis Model," *IEEE Transactions on Services Computing*, vol. 16, no. 2, pp. 830-844, 1 March-April 2023.

[53] F. Bi, T. He, Y. Xie and X. Luo, "Two-Stream Graph Convolutional Network-Incorporated Latent Feature Analysis," *IEEE Transactions on Services Computing*, vol. 16, no. 4, pp. 3027-3042, 1 July-Aug. 2023.

[54] T. He, Y. Liu, Y.S. Ong, X. Wu, and X. Luo, "Polarized message-passing in graph neural networks," *Artificial Intelligence*, vol. 331, pp. 104129, 2024.

[55] H.H. Zhou, T. He, Y. -S. Ong, G. Cong and Q. Chen, "Differentiable Clustering for Graph Attention," *IEEE Transactions on Knowledge and Data Engineering*, vol. 36, no. 8, pp. 3751-3764, Aug. 2024.

[56] J. Huang, Y. Lei, X. Li, "An adaptive generalized extended Kalman filter for real-time identification of structural systems, state and input based on sparse measurement," *Nonlinear Dynamics*, vol. 112, no. 2, pp. 1-24, Feb. 2024.

[57] T. Chen, S. Li, Y. Qiao and X. Luo, "A Robust and Efficient Ensemble of Diversified Evolutionary Computing Algorithms for Accurate Robot Calibration," *IEEE Transactions on Instrumentation and Measurement*, vol. 73, pp. 1-14, 2024, Art no. 7501814.

[58] F. Bi, T. He and X. Luo, "A Fast Nonnegative Autoencoder-Based Approach to Latent Feature Analysis on High-Dimensional and Incomplete Data," IEEE Transactions on Services Computing, vol. 17, no. 3, pp. 733-746, May-June 2024.

[59] F. Bi, T. He and X. Luo, "A Two-Stream Light Graph Convolution Network-based Latent Factor Model for Accurate Cloud Service QoS Estimation," in

[60] X. Fu, J. Sturtz, E. Alonso, R. Challoo and L. Qingge, "Parallel Trajectory Training of Recurrent Neural Network Controllers With Levenberg–Marquardt and Forward Accumulation Through Time in Closed-Loop Control Systems," *IEEE Transactions on Sustainable Computing*, vol. 9, no. 2, pp. 222-229, Mar 2024

[61] S. F. ur Rehman, Z. Mohamed, A. R. Husain, L. Ramli, M. A. Abbasi, W. Anjum and M. H. Shaheed, "Adaptive input shaper for payload swing control of a 5-DOF tower crane with parameter uncertainties and obstacle avoidance," *Automation in Construction*, vol. 154, no. 1, pp. 104963, Oct. 2023.

[62] T. He, Y.S. Ong, and L. Bai, "Learning conjoint attentions for graph neural nets," in *Proc. of Thirty-fifth Conference on Neural Information Processing Systems,* 2021, pp. 2641-2653.

[63] Y. Yuan, X. Luo and M. Zhou, "Adaptive Divergence-Based Non-Negative Latent Factor Analysis of High-Dimensional and Incomplete Matrices From Industrial Applications," *IEEE Transactions on Emerging Topics in Computational Intelligence*, vol. 8, no. 2, pp. 1209-1222, April 2024.

[64] L. Jin, S. Liang, X. Luo and M. Zhou, "Distributed and Time-Delayed -Winner-Take-All Network for Competitive Coordination of Multiple Robots," *IEEE Transactions on Cybernetics*, vol. 53, no. 1, pp. 641-652, Jan. 2023.

[65] L. Chen and X. Luo, "Tensor Distribution Regression Based on the 3D Conventional Neural Networks," *IEEE/CAA Journal of Automatica Sinica*, vol. 10, no. 7, pp. 1628-1630, Jul. 2023.

[66] D. Wu, Z. Li, Z. Yu, Y. He and X. Luo, "Robust Low-Rank Latent Feature Analysis for Spatiotemporal Signal Recovery," IEEE Transactions on Neural Networks and Learning Systems, DOI: 10.1109/TNNLS.2023.3339786.

[67] D. Wu, P. Zhang, Y. He and X. Luo, "MMLF: Multi-Metric Latent Feature Analysis for High-Dimensional and Incomplete Data," *IEEE Transactions on Services Computing*, vol. 17, no. 2, pp. 575-588, March-April 2024.

[68] Y. Yuan, R. Wang, G. Yuan and L. Xin, "An Adaptive Divergence-Based Non-Negative Latent Factor Model," *IEEE Transactions on Systems, Man, and Cybernetics: Systems*, vol. 53, no. 10, pp. 6475-6487, Oct. 2023.

[69] T. Chen, Y. Wang, H. Wen and J. Kang, "Autonomous assembly of multiple flexible spacecraft using RRT* algorithm and input shaping technique," *Nonlinear Dynamics*, vol. 111, no. 12, pp. 11223-11241, Apr. 2023.

[70] F. Bi, X. Luo, B. Shen, H. Dong and Z. Wang, "Proximal Alternating-Direction-Method-of-Multipliers-Incorporated Nonnegative Latent Factor Analysis," *IEEE/CAA Journal of Automatica Sinica*, vol. 10, no. 6, pp. 1388-1406, Jun. 2023.

[71] J. Chen, R. Wang, D. Wu and X. Luo, "A Differential Evolution-Enhanced Position-Transitional Approach to Latent Factor Analysis," *IEEE Transactions on Emerging Topics in Computational Intelligence*, vol. 7, no. 2, pp. 389-401, April 2023.

[72] W. Qin, X. Luo, S. Li and M. Zhou, "Parallel Adaptive Stochastic Gradient Descent Algorithms for Latent Factor Analysis of High-Dimensional and Incomplete Industrial Data," *IEEE Transactions on Automation Science and Engineering*, DOI: 10.1109/TASE.2023.3267609.

[73] A. Kou and X. Li, "Neural Network Intelligent Control Based on MPSO," *IEEE Access*, vol. 11, pp. 58565-58577, Jun. 2023.

[74] T. Xu, Y. Li, L. Yang, Q. Lin and W. Li, "Adaptive Control for Nonlinear State-Constrained Systems with Time-Varying Delays and Unknown Control Direction," in *Proc. of 2023 5th International Conference on Intelligent Control, Measurement and Signal Processing*, Chengdu, China, May. 2023, pp. 500-507.

[75] H. Chen, W. Xu, W. Guo and X. Sheng, "Variable Admittance Control Using Velocity-Curvature Pao Enhance Physical Human-Robot Interaction," *IEEE Robotics and Automation Letters*, vol. 9, no. 6, pp. 5054-5061, June 2024.

[76] X. Luo, J. Chen, Y. Yuan and Z. Wang, "Pseudo Gradient-Adjusted Particle Swarm Optimization for Accurate Adaptive Latent Factor Analysis," *IEEE Transactions on Systems, Man, and Cybernetics: Systems*, vol. 54, no. 4, pp. 2213-2226, Apr. 2024.

[77] X. Miao, L. Yang and H. Ouyang, "Artificial-neural-network-based optimal Smoother design for oscillation suppression control of underactuated overhead cranes with distributed mass beams," *Mechanical Systems and Signal Processing*, vol. 200, no. 1, pp. 110497, Oct. 2023.

[78] K. He, X. Zhang, S. Ren, and J. Sun, "Deep residual learning for image recognition," in *Proc. IEEE Conf. Comput. Vis. Pattern Recognit.*, 2016, pp. 770–778.

[79] A. Ravera et al., "Co-Design of a Controller and Its Digital Implementation: The MOBY-DIC2 Toolbox for Embedded Model Predictive Control," *IEEE Transactions on Control Systems Technology*, vol. 31, no. 6, pp. 2871-2878, Nov. 2023.

[80] Z. Lin and H. Wu, "Dynamical Representation Learning for Ethereum Transaction Network via Non-negative Adaptive Latent Factorization of Tensors," in *Proc. of 2021 International Conference on Cyber-Physical Social Intelligence*, Beijing, China, 2021, pp. 1-6.

[81] H. Wu, Y. Xia, and X. Luo, "Proportional Integral Derivative Incorporated Latent Factorization of Tensors for Large-Scale Dynamic Network Analysis," in *Proc. of 2021 China Automation Congress*, Beijing, China, 2021, pp. 2980-2984.

[82] Y. Shi, W. Sheng, S. Li, B. Li and X. Sun, "Neurodynamics for Equality-Constrained Time-Variant Nonlinear Optimization Using Discretization," *IEEE Transactions on Industrial Informatics*, vol. 20, no. 2, pp. 2354-2364, Feb. 2024.

[83] Y. Yuan, J. Li and X. Luo, "A Fuzzy PID-Incorporated Stochastic Gradient Descent Algorithm for Fast and Accurate Latent Factor Analysis," IEEE Transactions on Fuzzy Systems, DOI: 10.1109/TFUZZ.2024.3389733.

[84] Y. Shi, J. Wang, S. Li, B. Li and X. Sun, "Tracking Control of Cable-Driven Planar Robot Based on Discrete-Time Recurrent Neural Network With Immediate Discretization Method," *IEEE Transactions on Industrial Informatics*, vol. 19, no. 6, pp. 7414-7423, Jun. 2023.

[85] T. Tanabe and H. Kaneko, "Illusory Directional Sensation Induced by Asymmetric Vibrations Influences Sense of Agency and Velocity in Wrist Motions," *IEEE Transactions on Neural Systems and Rehabilitation Engineering*, DOI: 10.1109/TNSRE.2024.3393434.

[86] X. Luo, Z. Li, W. Yue and S. Li, "A Calibrator Fuzzy Ensemble for Highly-Accurate Robot Arm Calibration," *IEEE Transactions on Neural Networks and Learning Systems*, DOI: 10.1109/TNNLS.2024.3354080.